\documentclass{llncs}

\usepackage{graphicx}
\usepackage{tabularx}
\usepackage{amsmath}
\usepackage{amssymb}
\usepackage{microtype}
\usepackage{wrapfig}
\usepackage{enumitem}
\allowdisplaybreaks

\title{Modularity for Security-Sensitive Workflows 
\thanks{
This work has been partly supported by the EU under grant 317387 SECENTIS
(FP7-PEOPLE-2012-ITN).
}}

\author{Daniel R. dos Santos\inst{1,2,3}, Silvio Ranise\inst{1}, Serena E. Ponta\inst{2}}
\institute{$^1$ Fondazione Bruno Kessler (FBK)  $^2$SAP Labs France  $^3$University of Trento}

\begin{document}

\maketitle

\begin{abstract}
  An established trend in software engineering insists on using
  components (sometimes also called services or packages) to
  encapsulate a set of related functionalities or data.  By defining
  interfaces specifying what functionalities they provide or use,
  components can be combined with others to form more complex
  components.  In this way, IT systems can be designed by mostly
  re-using existing components and developing new ones to provide new
  functionalities.
  In this paper, we introduce a notion of component and a combination
  mechanism for an important class of software artifacts, called
  security-sensitive workflows.  These are business processes in which
  execution constraints on the tasks are complemented with
  authorization constraints (e.g., Separation of Duty) and
  authorization policies (constraining which users can execute which
  tasks). We show how well-known workflow execution patterns can be
  simulated by our combination mechanism and how authorization
  constraints can also be imposed across components.  Then, we
  demonstrate the usefulness of our notion of component by showing (i)
  the scalability of a technique for the synthesis of run-time
  monitors for security-sensitive workflows and (ii) the design of a
  plug-in for the re-use of workflows and related run-time
  monitors inside an editor for security-sensitive workflows.
\end{abstract}

\section{Introduction}
\label{sec:introduction}

Nowadays, business processes constantly strive to adapt to rapidly
evolving markets under continuous pressure of regulatory and
technological changes.  In this respect, the most frequent problem
faced by companies is the lack of automation when trying to
incorporate new business requirements into existing processes.  A
traditional approach to business process modeling frequently results
in large models that are difficult to change and maintain. This makes
it critical that business process models be modular and flexible, not
only for increased modeling agility at design-time but also for
greater robustness and flexibility of enacting processes at run-time
(see, e.g.,~\cite{markovic2007} for a discussion about this and related
problems).

The situation is further complicated when considering the class of
security-sensitive workflows~\cite{armando-ponta}, i.e.\ when tasks in
processes are executed under the responsibility of humans or software
agents acting on their behalf.  This means that, besides the usual
execution constraints (specified by causal relations among tasks),
security-sensitive workflows add authorization policies and
constraints, i.e.\ under which conditions users can execute tasks.
Authorization policies are usually specified by using some variant of
the Role Based Access Control (RBAC) model, see,
e.g.,~\cite{wainer2007}, while authorization constraints restrict
which users can execute some set of tasks in a given workflow
instance; an example is the Separation of Duties (SoD) constraint
requiring two tasks to be executed by distinct users.  Since
authorization policies and constraints may prevent the successful
termination of the workflow (i.e.\ not all tasks can be executed), it
is crucial to be able to solve at design-time, the Workflow 
Satisfiability Problem (WSP)~\cite{crampton}, i.e.\ establishing if all 
tasks in the workflow can be executed satisfying the authorization policy 
without violating any authorization constraint, and at run-time, a 
variant of the WSP requiring the synthesis of a monitor capable of 
granting the request of a user to execute a task if this does not prevent 
the successful termination of the workflow instance (see, 
e.g.,~\cite{basin2012,bertolissi2015}).
The combination of the need for modularity and flexibility with that
for developing efficient techniques to solve the WSP and its run-time
variant gives rise to new fundamental questions, such as
\begin{description}
\item[Q1:] how can we specify security-sensitive workflow components,
  i.e.\ business processes equipped with interfaces defining their
  inputs and outputs together with their dependencies (a component
  declares the services it provides and those that it depends upon)?  
\item[Q2:] how can we ``glue together'' components into a more complex
  one that can again be combined with others if necessary?
\item[Q3:] how can we solve the WSP and synthesize run-time monitors
  for security-sensitive workflow components that can be modularly
  re-used to solve the WSP and synthesize a run-time monitor for their
  combination?
\end{description}
In this paper, we provide answers to the three questions above by
making the following contributions:
\begin{description}
\item[A1:] we introduce the notion of security-sensitive workflow
  component (Section~\ref{sec:reusable}) as a symbolic transition
  system extended with a suitable notion of interface,
\item[A2:] we define how components can be ``glued together''
  (Section~\ref{sec:glue}) by specifying how execution and
  authorization constraints of components become related,
\item[A3:] we describe how run-time monitors solving the WSP of
  security-sensitive workflow components can be modularly reused
  (Section~\ref{sec:applications}) to build one solving the WSP of their
  combination.
\end{description}
We show the adequacy of \textbf{A1} and \textbf{A2} by showing how a
typical security-sensitive workflow can be specified as a composition
of components (Section~\ref{sec:reusable}).  (\textbf{A2} is further
elaborated in Appendix~\ref{app:patterns} by demonstrating how the
main composition patterns for workflows, such as those in~\cite{YAWL},
can be simulated by our notion of gluing.)

Another contribution of the paper is an investigation of how our
proposal can be exploited in an industrial setting
(Section~\ref{sec:applications}).  In particular, we consider two
main issues.  First, we show how splitting into several modules large
security-sensitive workflows, by using \textbf{A1} and \textbf{A2},
allows for the synthesis of run-time monitors to scale up, by using
\textbf{A3}.  Second, we sketch the architecture of a tool for the 
creation of security-sensitive workflows which maintains a library of 
components together with their run-time monitors.  This holds the promise 
to help workflow designers in their quest for adapting processes to rapidly 
evolving requirements.

\section{Security-sensitive Workflow Components}
\label{sec:reusable}
We introduce a refinement of the notion of symbolic transition system
in~\cite{bertolissi2015} which, associated to a suitable notion of
interface, constitutes a (symbolic) security-sensitive component.  We
motivate the utility of this notion by means of an example.
\begin{example}
 \label{ex:trw}
\begin{figure}[t]
  \centering	
  \includegraphics[scale=0.42]{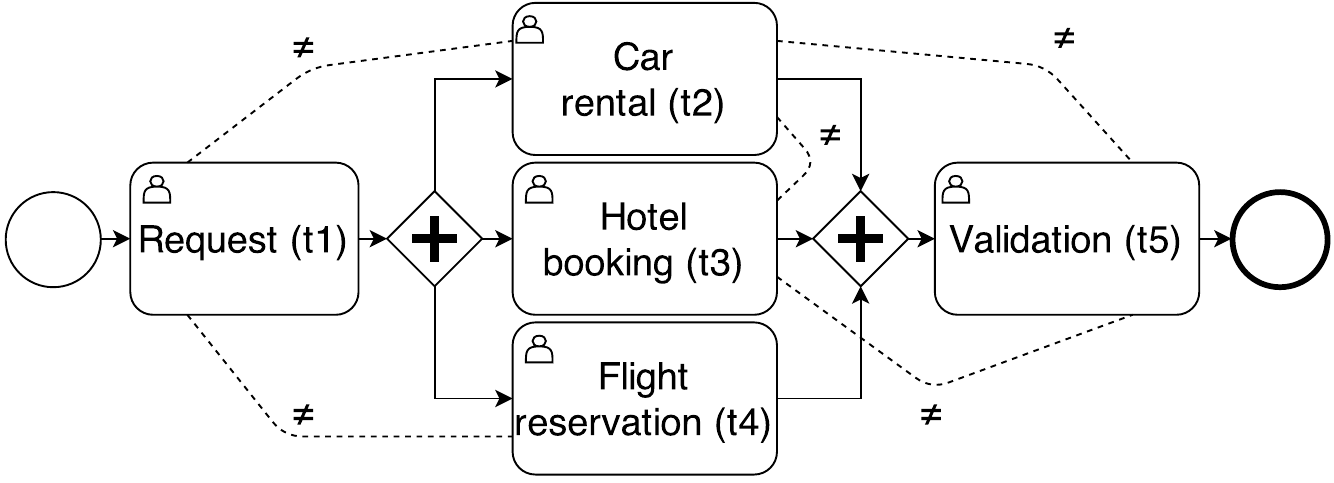}
  \includegraphics[scale=0.42]{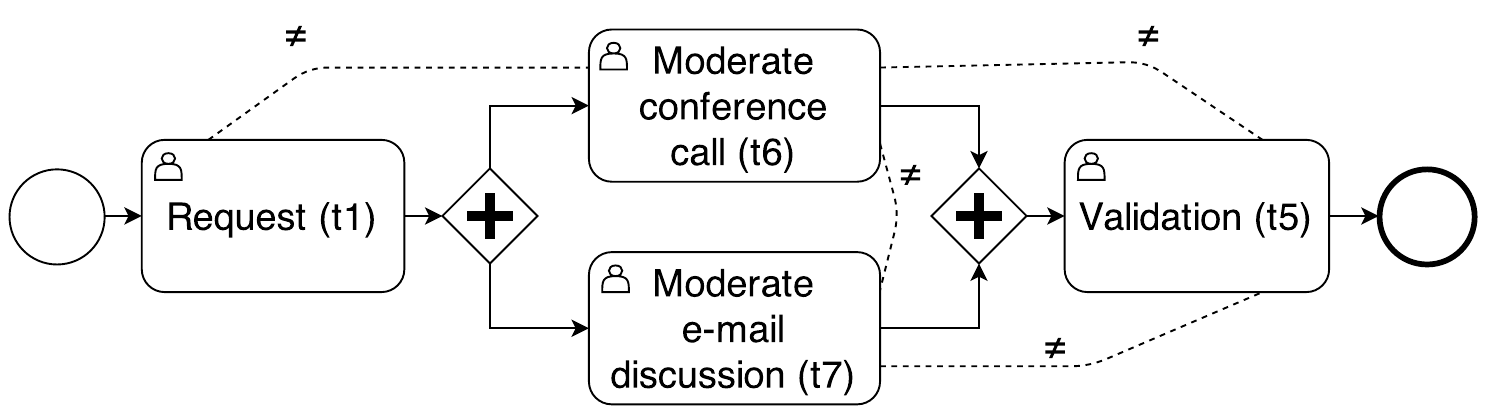}	
  \caption{\label{fig:wf-ex}TRW (left) and MDW (right) in extended BPM
    notation}
\end{figure}
 Figure~\ref{fig:wf-ex} shows two workflows in BPM Notation
 (BPMN)~\cite{omg2011}.  Each workflow contains two circles, the one
 on the left represents the start event (triggering the execution of
 the workflow), whereas that on the right the end event (terminating
 the execution of the workflow), tasks are depicted by labeled boxes,
 the constraints on the execution of tasks are shown as solid arrows
 (for sequence flows) and diamonds labeled by $+$ (for parallel
 flows), the fact that a task must be executed under the
 responsibility of a user is indicated by the man icon inside a box,
 and the SoD constraints as dashed lines labeled by $\neq$.

 The workflow on the left is the Trip Request Workflow (TRW) whose
 goal is that of requesting trips for employees in an organization. It
 is composed of five tasks: Request ($t1$), Car rental ($t2$), Hotel
 booking ($t3$), Flight reservation ($t4$), and Validation ($t5$).
 Five Separation of Duty (SoD) constraints must be enforced, i.e.\ the
 tasks in the pairs $(t1,t2)$, $(t1,t4)$, $(t2,t3)$, $(t2,t5)$, and
 $(t3,t5)$ must be executed by distinct users in any sequence of task
 executions of the TRW.   

 The workflow on the right is the Moderate Discussion Workflow (MDW)
 whose goal is to organize a discussion and voting process in an 
 organization. It is composed of four tasks: Request ($t1$), 
 Moderate Conference Call ($t7$), Moderate e-mail Discussion ($t7$), 
 and Validation ($t5$).  Four SoD constraints must be enforced: 
 $(t1,t6)$, $(t6,t5)$, $(t6,t7)$, and $(t7,t5)$.

 In both workflows, each task is executed under the responsibility of
 a user who has the right to execute it according to some
 authorization policy, which---for the sake of brevity---we leave
 unspecified.  

 Notice that tasks $t1$ and $t5$ in Figure~\ref{fig:wf-ex} are the
 same in both TRW and MDW.  The goal of this paper is to develop a
 notion of security-sensitive component such that tasks $t1$ and $t5$
 can be modularly reused in the specifications of both workflows so
 that only the specification of the parallel execution of tasks $t2$,
 $t3$, and $t4$ for the TRW and $t6$ and $t7$ for the MDW must be 
 developed from scratch in the two cases.  Additionally, we want that 
 run-time monitors for the various components can also be modularly 
 reused.

 Indeed, the simplicity of the TRW and MDW spoils the advantages of a
 modular approach; the small dimension of the workflows allows us to
 keep the paper to a reasonable size.  However, for large
 workflows---as we will see below in
 Section~\ref{sec:applications}---the advantages are substantial.  To
 give an intuition of this, imagine to replace the tasks reused in
 both workflows, i.e.\ $t1$ and $t5$, with complex workflows: reusing
 their specifications and being able to synthesize run-time monitors
 for them, that can be used for larger workflows in which they are 
 plugged, becomes much more interesting. \qed
\end{example}
A \emph{(symbolic) security-sensitive component} is a pair
$(S,\mathit{Int})$ where $S$ is a {(symbolic) security-sensitive
  transition system} and $\mathit{Int}$ is its {interface}.

\noindent \textbf{Security-sensitive transition system}. Since the
semantics of BPMN can be given by means of (extensions of) Petri nets
(see, e.g.,~\cite{YAWL}) and the latter can be represented as symbolic
state transition systems (see, e.g.,~\cite{sankaranarayanan2003}), $S$
is the symbolic transition system that can be associated to security
sensitive workflows specified in BPMN as those in
Figure~\ref{fig:wf-ex}.  
A \emph{(symbolic) security-sensitive
  transition system} $S$ is a tuple of the form
$((P,D,A,H,C),\mathit{Tr},B)$ where $P\cup D\cup A\cup H\cup C$ are
the state variables, $\mathit{Tr}$ is a set of transitions, and $B$ is
a set of constraints on the state variables in $C$.  The finite set
$P$ contains Boolean variables representing the places of the Petri
net associated to a BPMN specification of the security-sensitive
workflow and $D$ is a finite set of Boolean variables representing the
fact that a task has been executed or not; $P\cup D$ are called
\emph{execution constraint variables}.  The finite set $A$ contains
interface predicates to the authorization policy, $H$ is a set of
predicates recording which users have executed which tasks, and $C$ is
a set of interface predicates to the authorization constraints; $A\cup
H\cup C$ are called \emph{authorization constraint variables}.  The
set $\mathit{Tr}$ contains the \emph{transitions} (or \emph{events})
of the form
\begin{eqnarray}
  \label{eq:p-event}
  t(u) : 
  \mathit{en}_{\mbox{\scriptsize EC}}(P,D)\wedge 
  \mathit{en}_{\mbox{\scriptsize Auth}}(A,C)
  \to 
  \mathit{act}_{\mbox{\scriptsize EC}}(P,D) || 
  \mathit{act}_{\mbox{\scriptsize Auth}}(H)
\end{eqnarray}
where $t$ is the name of a task taken from a finite set, $u$ is a
variable ranging over a set $U$ of users,
$\mathit{en}_{\mbox{\scriptsize EC}}(P,D)$ is a predicate on $P\cup D$
(called the \emph{enabling condition for the execution constraint}),
$\mathit{en}_{\mbox{\scriptsize Auth}}(A,C)$ is a predicate on $\{v(u)
| v\in A\cup C\}$ (called the \emph{enabling condition for the
  authorization constraint}), $\mathit{act}_{\mbox{\scriptsize
    EC}}(P,D)$ contains parallel assignments of the form $v:=b$ where
$v\in P\cup D$ and $b$ is a Boolean value (called the \emph{update of
  the execution constraint} of the security sensitive workflow), and
$\mathit{act}_{\mbox{\scriptsize Auth}}(H)$ contains parallel
assignments of the form $v(u):=b$ where $v\in H$ and $b$ is a Boolean
value (called the \emph{update of the authorization history} of the
security sensitive workflow).\footnote{The assignment $v(u):=b$
  leaves unchanged the value returned by $v$ for any $u'$ distinct
  from $u$.  In other words, after the assignment, the value of $v$
  can be expressed as follows: $\lambda
  x.\mathit{if}~x=u~\mathit{then}~b~\mathit{else}~v(x)$.} Finally, the
finite set $B$ contains \emph{always constraints} of the form
\begin{eqnarray}
  \label{eq:always-constr}
  \forall u.v(u) \Leftrightarrow \mathit{hst},
\end{eqnarray} 
where $u$ is a variable ranging over users, $v$ is a variable in $C$,
and $\mathit{hst}$ is a Boolean combination of atoms of the form
$w(u)$ with $w\in H$.

\noindent \textbf{Interface of a security-sensitive
  component}. \label{int:beginning} The interface $\mathit{Int}$ of a
symbolic security-sensitive component $(S,\mathit{Int})$ is a tuple of
the form $(A,P^i,P^o$, $H^o,C^i)$ where
\begin{itemize}
  \label{page:def-comp}
\item $P^i\subseteq P$ and each $p^i\in P^i$ is such that $p^i:=T$
  does not occur in the parallel assignments of an event of the form
  (\ref{eq:p-event}) in $\mathit{Tr}$,
\item $P^o\subseteq P$ and each $p^o\in P^o$ is such that $p^o:=T$
  occurs in the parallel assignments of an event of the form
  (\ref{eq:p-event}) in $\mathit{Tr}$ whereas $p^o:=F$ does not,
\item $H^o\subseteq H$, $C^i\subseteq C$, and
\item only the variables in $(C\setminus C^i)\cup H^o$ can occur in a
  symbolic always constraint of $B$.
\end{itemize}
When $P^i$, $P^o$, $H^o$, and $C^i$ are all empty, the component
$(S,\mathit{Int})$ can only be interfaced with an authorization policy
via the interface variables in $A$.  The state variables in $D$ are
only used internally, to indicate that a task has been or has not been
executed; thus, none of them is exposed in the interface
$\mathit{Int}$.  The variables in $P$, $H$, and $C$ are local to $S$
but some of them can be exposed in the interface in order to enable
the combination of $S$ with other components in a way which will be
described below (Section~\ref{sec:glue}).  The super-scripts $i$ and
$o$ stand for input and output, respectively.  The requirement that
variables in $P^i$ are not assigned the value $T$(rue) by any
transition of the component allows their values to be determined by
those in another component.  Dually, the requirement that variables in
$P^o$ can only be assigned the value $T$(rue) by any transition of the
component allows them to determine the values of variables in another
component.  Similarly to the values of the variables in $P^i$, those
of the variables in $C^i$ are fixed when combining the module with
another; this is the reason for which only the variables in
$C\setminus C^i$ can occur in the always constraints of the component.

\begin{example}
\label{ex:trw-ex}
We now illustrate the notion of security-sensitive component by
considering the workflows in Figure~\ref{fig:wf-ex}.
\begin{figure}[t]
 \centering	
 \includegraphics[scale=0.75]{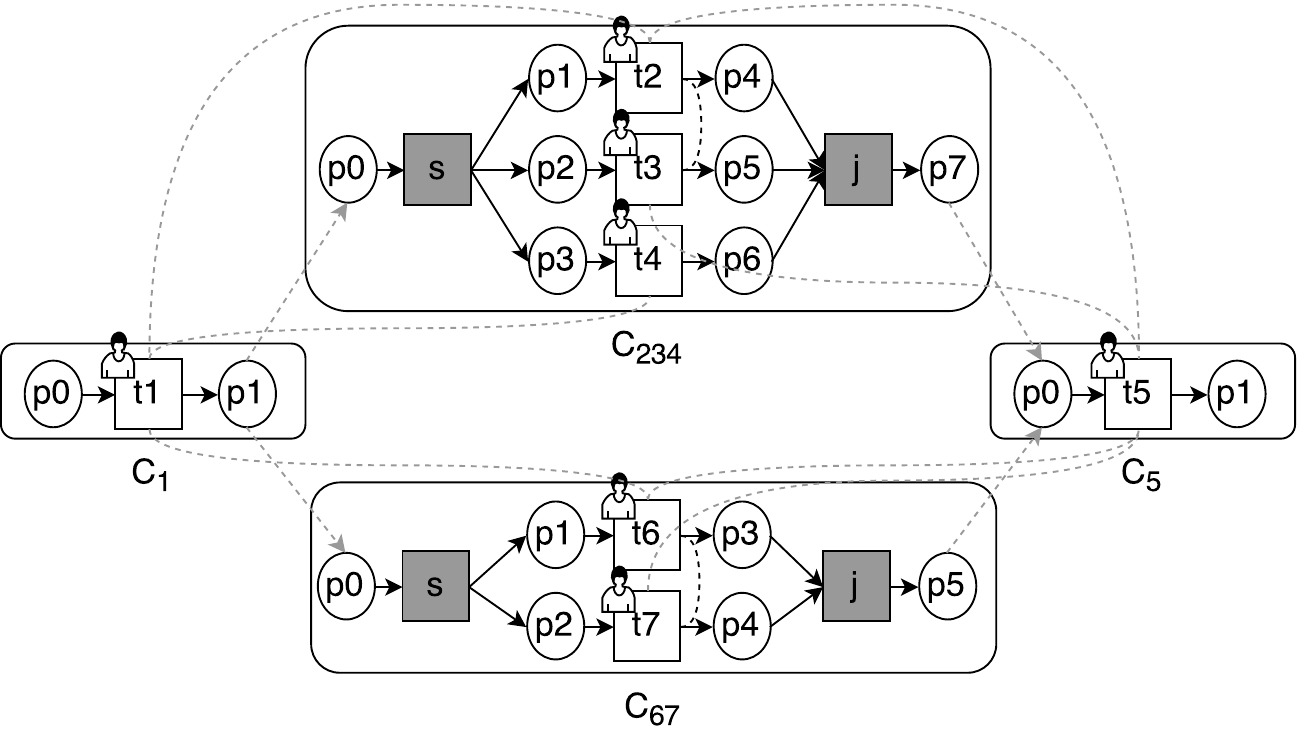}	
 \caption{\label{fig:wf-pn}TRW and MDW as combinations of security-sensitive components}
\end{figure}
As said in Example~\ref{ex:trw}, we want to reuse tasks $t1$ and $t5$
in both TRW and MDW.  For this, we split the specification of each
workflow in four components $C_1$, $C_{234}$, $C_{67}$, and $C_{5}$ as
shown in Figure~\ref{fig:wf-pn}, where the sequential composition of
$C_1$, $C_{234}$, and $C_{5}$ yields TRW and that of $C_1$, $C_{67}$,
and $C_{5}$ gives MDW.  The figure shows the extended Petri nets
representing the four components and how they are connected: circles
represent places, rectangles with a man icon transitions to be
executed under the responsibility of users, rectangles without the
icon transitions not needing human intervention, (black) dashed lines
represent SoD constraints between tasks belonging to the same
component, (gray) dashed lines SoD constraints between tasks belonging
to distinct components, (black) solid arrows the control flow in the
same component, and (gray) dashed arrows the control flow between two
components. Note that the control flow between two components is outside
  of the semantics of extended Petri nets. For example, a token in
place $p0$ of $C_1$ goes to $p1$ of $C_1$ after the execution of $t1$
and, at the same time a token is put in place $p1$ of $C_{234}$
because of the (gray) dashed arrow from $p1$ in $C_1$ to $p0$ in
$C_{234}$ representing an inter execution constraint.  When the token
is in $p0$, the system executes the split transition $s$ in $C_{234}$
that removes the token from $p0$ and puts one in $p1$, $p2$, and $p3$
so that $t2$, $t3$, and $t4$ in $C_{234}$ become enabled.  Notice that
the execution of $t2$ is constrained by a SoD constraint from task
$t1$ in component $C_1$ (dashed arrow between $t1$ in $C_1$ and $t2$
in $C_{234}$): this means that the user who has executed $t1$ in $C_1$
cannot execute also $t2$ in $C_{234}$.

We now show how to formalize the components depicted in
Figure~\ref{fig:wf-pn} by defining $\mathcal{C}_{1}=(S_{1},
\mathit{Int}_{1})$, $\mathcal{C}_{5}=(S_{5}, \mathit{Int}_{5})$,
$\mathcal{C}_{234}=(S_{234}, \mathit{Int}_{234})$, and
$\mathcal{C}_{67}=(S_{67}, \mathit{Int}_{67})$ where
$S_y=((P_y,D_y,A_y,H_y,C_y),\mathit{Tr}_y,B_y)$, and $\mathit{Int}_y=
(A_y,P_y^i,P_y^o, H_y^o,C_y^i)$ for $y=1,5,234,67$.  For components
$\mathcal{C}_1$ and $\mathcal{C}_5$, we set
\begin{eqnarray*}
  \begin{array}{l}
  \small
  P_y := \{p0_{y},p1_{y}\} \,\, D_y:= \{d_{ty}\} \,\, A_y:= \{a_{ty}\} \,\, 
  H_y:= \{h_{ty}\} \,\, C_y:= \{c^i_{ty}\}\,\, B_y:=\emptyset \\ 
  P_y^i:=\{p0_{y}\}\,\, P_y^o:=\{p1_{y}\}\,\, H_y^o:=\{h_{ty}\}\, .
  \end{array}
\end{eqnarray*}
for $y=1,5$, and take
\begin{eqnarray*}
  \begin{array}{l}
  \small
   \mathit{Tr}_1 := \{ t1(u) : p0_{1}\wedge \neg d_{t1} \wedge a_{t1}(u) \to p0_{1},p1_{1},d_{t1},h_{t1}(u):=F,T,T,T\} \\
  \mathit{Tr}_5 := \{ t5(u) : p0_{5}\wedge \neg d_{t5} \wedge a_{t5}(u)\wedge c^i_{t5}(u) \to p0_{5},p1_{5},d_{t5},h_{t5}(u):=F,T,T,T\} \\
  C_1^i:=\emptyset \,\,\,\, C_5^i:=\{c^i_{t5}\}\, .
  \end{array}
\end{eqnarray*}
According to the transition in $\mathit{Tr}_1$, task $t1$ is enabled
when there is a token in place $p0_{1}$ (place $p0$ of component
$\mathcal{C}_1$ in Figure~\ref{fig:wf-pn}), $t1$ has not been already
executed ($\neg d_{t1}$) and there exists a user $u$ capable of
executing $t1$ ($a_{t1}(u)$).  The effect of executing such a
transition is to move the token from $p0_{1}$ to $p1_{1}$ (places $p0$
and $p1$ of component $\mathcal{C}_1$ in Figure~\ref{fig:wf-pn},
respectively), set $d_{t1}$ to true meaning that $t1$ has been
executed, and recording that $t1$ has been executed by $u$.  The
interface of each component is the following: $p0_{y}$ is the input
place, $p1_{y}$ is the output place, and the history variable $h_{ty}$
can be used to constrain the execution of tasks in other components
(for instance of $t2$ in the TRW as $t1$ and $t2$ are involved in a
SoD, shown by the gray dashed line between the two tasks in
Figure~\ref{fig:wf-pn}).  Notice that the execution of task $t1$
cannot be constrained by the execution of tasks in other components
(thus $C_1^i:=\emptyset$) since $t1$ is always executed before all
other tasks and cannot possibly be influenced by their execution.  The
definition of the transition in $\mathit{Tr}_5$ is similar to that in
$\mathit{Tr}_1$ except for the fact that the execution of task $t5$
can be constrained by the execution of tasks in other components (thus
$C_5^i:=\{c^i_{t5}\}$) since $t5$ is always executed after all other
tasks and can be influenced by their execution.  In particular,
$c^i_{t5}$ will be defined so as to satisfy the SoD constraints
between $t5$ and $t2$ or $t3$ for TRW and $t6$ or $t7$ for MDW.  
For component $\mathcal{C}_{234}$, we set
\begin{eqnarray*}
  \begin{array}{l}
  \small
  P_{234} := \{py_{234} | y=0, ..., 7\} \,\, 
  D_{234}:= \{s_{234},j_{234},d_{ty}| y=2,3,4\} \\
  A_{234}:= \{a_{ty}| y=2,3,4\} \,\,
  H_{234}:= \{h_{ty}| y=2,3,4 \} \,\, 
  C_{234}:= \{c_{t2},c_{t3},c^i_{ty} | y=2,3,4 \} \\
  B_{234}:=\{\forall u. c_{t2}(u)\Leftrightarrow \neg h_{t3}(u),
           \forall u. c_{t3}(u)\Leftrightarrow \neg h_{t2}(u)\} \\ 
  \mathit{Tr}_{234} := \left\{ 
  \begin{array}{lcl}
    s_{234} &:& p0_{234}\wedge\neg d_s \to p0_{234}, p1_{234}, p2_{234},p3_{234}, d_{s_{234}}:= F,T,T,T,T \\
    t2(u) &:& p1_{234}\wedge \neg d_{t2} \wedge a_{t2}(u)\wedge c_{t2}(u)\wedge c^i_{t2}(u) \\ 
    &&\to p1_{234},p4_{234},d_{t2},h_{t2}(u):=F,T,T,T \\
    t3(u) &:& p2_{234}\wedge \neg d_{t3} \wedge a_{t3}(u)\wedge c_{t3}(u)\wedge c^i_{t3}(u) \\
    &&\to p2_{234},p5_{234},d_{t3},h_{t3}(u):=F,T,T,T \\
    t4(u) &:& p3_{234}\wedge \neg d_{t4} \wedge a_{t4}(u)\wedge c^i_{t4}(u) \\
    &&\to p3_{234},p6_{234},d_{t4},h_{t4}(u):=F,T,T,T \\
    j_{234} &:& p4_{234}\wedge p5_{234}\wedge p6_{234}\wedge \neg d_j \\
    &&\to p4_{234}, p5_{234},p6_{234}, p7_{234}, d_{j_{234}}:= F,F,F,T,T
  \end{array}\right\} \\
  P_{234}^i:=\{p0_{234}\}\,\, 
  P_{234}^o:=\{p7_{234}\}\,\, 
  H_{234}^o:=\{h_{t2},h_{t3}\}\,\, 
  C_{234}^i:=\{c^i_{t2},c^i_{t4} \}.
  \end{array}
\end{eqnarray*}
Transitions $s_{234}$ and $j_{234}$ (corresponding to the rectangles
labeled $s$ and $j$ of component $\mathcal{C}_{234}$ in
Figure~\ref{fig:wf-pn}) model the parallel composition of tasks $t2$,
$t3$, and $t4$ in TRW and MDW (cf.\ the parallel flows depicted as
diamonds labeled with $+$ in Figure~\ref{fig:wf-ex}).  Since no human
intervention is needed, the enabling conditions for the authorization
constraint of both transitions are omitted.  Tasks $t2$ and $t3$ are
involved in a SoD constraint (cf.\ the dashed lines labeled by $\neq$
between $t2$ and $t3$ in Figure~\ref{fig:wf-pn}).  For this reason,
their enabling conditions contain $c_{t2}(u)$ and $c_{t3}(u)$ which
are defined in $B_{234}$ so as to prevent the execution of $t2$ and
$t3$ by the same users: to execute $t3$ ($t2$, resp.), user $u$ must
be such that $\neg h_{t2}(u)$ ($\neg h_{t3}(u)$, resp.), i.e.\ $u$
should have not executed $t2$ ($t3$, resp.).  Transitions $t2$, $t3$,
and $t4$ in $\mathit{Tr}_{234}$ have enabling conditions that contain
$c^i_{t2}(u)$, $c^i_{t3}(u)$, and $c^i_{t4}(u)$ which will be defined
so as to satisfy the SoD constraints in which the tasks are involved
(cf.\ the gray dashed lines across the rectangles in
Figure~\ref{fig:wf-pn}).  The definition of component
$\mathcal{C}_{67}$ is quite similar (albeit simpler) to that of
$\mathcal{C}_{234}$: 
\begin{eqnarray*}
  \begin{array}{l}
  \small
  P_{67} := \{py_{67} | y=0, ..., 5\} \,\, 
  D_{67}:= \{s_{67},j_{67},d_{ty}| y=6,7\} \,\, 
  A_{67}:= \{a_{ty} | y=6,7\} \\
  H_{67}:= \{h_{ty}| y=6,7 \} \,\, 
  C_{67}:= \{c_{ty},c^i_{ty} | y=6,7 \} \\
  B_{67}:=\{\forall u. c_{t6}(u)\Leftrightarrow \neg h_{t7}(u),
         \forall u. c_{t7}(u)\Leftrightarrow \neg h_{t6}(u)\} \\ 
  \mathit{Tr}_{67} := \left\{ 
  \begin{array}{lcl}
    s_{67} &:& p0_{67}\wedge\neg d_s \to p0_{67}, p1_{67}, p2_{67}, d_{s_{67}}:= F,T,T,T \\
    t6(u) &:& p1_{67}\wedge \neg d_{t6} \wedge a_{t6}(u)\wedge c_{t6}(u)\wedge c^i_{t6}(u) \\
    &&\to p1_{67},p3_{67},d_{t6},h_{t6}(u):=F,T,T,T \\
    t7(u) &:& p2_{67}\wedge \neg d_{t7} \wedge a_{t7}(u)\wedge c_{t7}(u)\wedge c^i_{t7}(u) \\
    &&\to p2_{67},p4_{67},d_{t7},h_{t7}(u):=F,T,T,T \\
    j_{67} &:& p3_{67}\wedge p4_{67}\wedge \neg d_j \to p3_{67}, p4_{67},p5_{67}, d_{j_{67}}:= F,F,T,T
  \end{array}\right\} \\
  P_{67}^i:=\{p0_{67}\}\,\, 
  P_{67}^o:=\{p5_{67}\}\,\, 
  H_{67}^o:=\{h_{t6},h_{t7}\}\,\, 
  C_{67}^i:=\{c^i_{t6}\}.
  \end{array}
\end{eqnarray*}
Section~\ref{sec:glue} below explains how components $\mathcal{C}_1$,
$\mathcal{C}_{234}$, $\mathcal{C}_{67}$, and $\mathcal{C}_{5}$ can be
``glued together'' to build TRW and MDW.  \qed
\end{example}
\noindent \textbf{Semantics of a security-sensitive component}.
The notion of symbolic security-sensitive transition system introduced
here is equivalent to that in~\cite{bertolissi2015}; the only
difference being the presence of the authorization constraint
variables in $C$ together with the always constraints in $B$.  It is
easy to see that, given a transition system
$((P,D,A,H,C),\mathit{Tr},B)$, it is always possible to eliminate the
variables in $C$ occurring in $B$ from the conditions of transitions
in $\mathit{Tr}$ by using (\ref{eq:always-constr}): it is sufficient
to replace each occurrence of $v(u)$ with $\mathit{hst}$.  Let
$[[\mathit{tr}]]_B$ denote the transition obtained from $\mathit{tr}$
by exhaustively replacing the variables in $C$ that also occur in $B$
as explained above.  Since no variable in $C$ may occur in the update
of a transition and in the enabling condition for the execution
constraint of a transition, by abuse of notation, we apply the
operator $[[\cdot]]_B$ to the enabling condition for the authorization
constraint of $\mathit{tr}$.  The substitution process eventually
terminates since in $\mathit{hst}$ there is no occurrence of variables
in the finite set $C$, only the variables in $H$ may occur.
The possibility of eliminating the variables in $C$ allows us to give
the semantics of the class of (symbolic) security-sensitive transition
systems considered here by using the notion of weakest liberal
precondition (wlp)~\cite{wlp} as done in~\cite{bertolissi2015}.  The
intuition is that computing a wlp with respect to the transitions in
$\mathit{Tr}$ and the always constraints in $B$ is equivalent to
computing that with respect to $[[\mathit{Tr}]]_B$.  Formally, we
define
\begin{eqnarray}
  \label{eq:wlp-def}
  \mathsf{wlp}(\mathit{Tr}, B, K) &:=& 
   \bigvee_{\mathit{tr}\in \mathit{Tr}} 
    (\mathit{en}_{\mbox{\scriptsize EC}}\wedge 
     [[\mathit{en}_{\mbox{\scriptsize Auth}}]]_B \wedge 
     K[\mathit{act}_{\mbox{\scriptsize EC}}||\mathit{act}_{\mbox{\scriptsize Auth}}])
\end{eqnarray}
where $B$ is a set of always constraints, $\mathit{tr}$ is of the form
(\ref{eq:p-event}), $K$ is a predicate over $P\cup D\cup A\cup C$, and
$K[\mathit{act}_{\mbox{\scriptsize
      EC}}||\mathit{act}_{\mbox{\scriptsize Auth}}]$ denotes the
predicate obtained from $K$ by  substituting 
\begin{itemize}
\item each variable $v\in P\cup D$ with the value $b$ when the
  assignment $v:=b$ is in $\mathit{act}_{\mbox{\scriptsize EC}}$ and
\item each variable $v\in H$ with $\lambda
  x.\mathit{if}~x=u~\mathit{then}~b~\mathit{else}~v(x)$ when $v(x):=b$
  is in $\mathit{act}_{\mbox{\scriptsize Auth}}$ for $b$ a Boolean
  value.
\end{itemize}
It is easy to show that $\mathsf{wlp}(\mathit{Tr}, B, K)$ is
$\bigvee_{\mathit{tr}\in \mathit{Tr}}\mathsf{wlp}(\mathit{tr}, B, K)$.
When $\mathit{Tr}$ is a singleton containing one symbolic transition
$\mathit{tr}$, we write $\mathsf{wlp}(\mathit{tr}, B, K)$ instead of
$\mathsf{wlp}(\{\mathit{tr}\}, B, K)$.  

A \emph{symbolic behavior} of a security-sensitive transition system
$S=((P,D,A,H$, $C),\mathit{Tr},B)$ is a sequence of the form $K_0
\stackrel{\mathit{tr}_0}{\longrightarrow} K_1
\stackrel{\mathit{tr}_1}{\longrightarrow}\cdots
\stackrel{\mathit{tr}_{n-1}}{\longrightarrow} K_n$ where $K_i$ is a
predicate over $P\cup D\cup A\cup C$ and $\mathit{tr}_i$ is a symbolic
transition such that $K_i$ is logically equivalent to
$\mathsf{wlp}(\mathit{tr}_i, B, K_{i+1})$ for $i=0, ..., n-1$.  The
semantics of the security-sensitive transition system $S$ is the set
of all possible symbolic behaviors.
The semantics of a security-sensitive component $(S,\mathit{Int})$ is
the set of all possible symbolic behaviors of the security-sensitive
transition system $S$.

\begin{example}
  We consider component $\mathcal{C}_5$ (cf.\ Example~\ref{ex:trw-ex})
  and compute the wlp with respect to $t5(u)$ (in the set
  $\mathit{Tr}_5$ of transitions) for the following predicate
  $\neg p0_5\wedge p1_5 \wedge d_{t5}$ 
  characterizing the set of final states of $\mathcal{C}_5$,
  i.e.\ those states in which  task $t5$ has been executed and
  there is just one token in place $p1_5$.  By using definition
  (\ref{eq:wlp-def}) above, we obtain 
  $p0_5\wedge \neg d_{t5}\wedge a_{t5}(u)\wedge c^i_{t5}(u)$,
  which identifies those states in which there is a token in place
  $p0_1$, task $t1$ has not yet been executed, and user $u$ has the
  right to execute $t1$ and authorization constraints imposed by other
  components are satisfied (e.g., the SoD constraint between $t5$ and
  $t2$ in $\mathcal{C}_{234}$ for the TRW). \qed
\end{example}

\section{Gluing together Security-Sensitive Components}
\label{sec:glue}
We now show how components can be combined together in order to build
other, more complex, components.  For $l=1,2$, let $(S_l,
\mathit{Int}_l)$ be a symbolic security-sensitive component where
$\mathit{Int}_l=(A,P_l^i,P_l^o, H_l^o,C_l^i)$ and
$S_l=((P_l,D_l,A_l,H_l,$ $C_l),\mathit{Tr}_l,B_l)$ is such that $P_1$
and $P_2$, $D_1$ and $D_2$, $A_1$ and $A_2$, $H_1$ and $H_2$, 
$C_1$ and $C_2$ 
are pairwise disjoint sets.  Furthermore, let $G=G_{\mbox{\scriptsize
    EC}}\cup G_{\mbox{\scriptsize Auth}}$ be a set of \emph{gluing
  assertions over $\mathit{Int}_1$ and $\mathit{Int}_2$}, where
\begin{itemize}
\item $G_{\mbox{\scriptsize EC}}$ is a set of formulae of the form
  $p^i\Leftrightarrow p^o$ for $p^i\in P_k^i$ and $p^o\in P_j^o$,
  called \emph{inter execution constraints}, and
\item $G_{\mbox{\scriptsize Auth}}$ is a set of always constraints in
  which only the variables in $C_k^{i}\cup H_j^{o}$ may occur,
\end{itemize}
for $k,j=1,2$ and $k\neq j$.  Intuitively, the gluing assertions in
$G$ specify inter component constraints; those in
$G_{\mbox{\scriptsize EC}}$ how the control flow is passed from one
component to another whereas those in $G_{\mbox{\scriptsize Auth}}$
authorization constraints across components, i.e.\ how the fact that a
task in a component is executed by a certain user constrains the
execution of a task in another component by a sub-set of the users
entitled to do so.

The \emph{symbolic security-sensitive component $(S,\mathit{Int})$
  obtained by gluing $(S_1,\mathit{Int}_1)$ and $(S_2,\mathit{Int}_2)$
  together with $G$}, in symbols
$(S,\mathit{Int})=(S_1,\mathit{Int}_1)\oplus_G (S_2,\mathit{Int}_2)$,
is defined as $S=((P,D,A,H, C),\mathit{Tr},B)$ and
$\mathit{Int}=(A,P^i,P^o, H^o,C^i)$, where
\begin{itemize}
\item $P=P_1\cup P_2$, $D=D_1\cup D_2$, $A=A_1\cup A_2$, $H=H_1\cup
  H_2$, $C=C_1\cup C_2$,
\item $\mathit{Tr}:=[\mathit{Tr}_1]_{G_{\mbox{\scriptsize EC}}}\cup
  [\mathit{Tr}_2]_{G_{\mbox{\scriptsize EC}}}$ where
  $[\mathit{Tr}_j]_{G_{\mbox{\scriptsize
        EC}}}:=\{[\mathit{tr}_j]_{G_{\mbox{\scriptsize EC}}}|
  \mathit{tr}_j\in \mathit{Tr}_j\}$ and
  $[\mathit{tr}_j]_{G_{\mbox{\scriptsize EC}}}$ is obtained from
  $\mathit{tr}_j$ by adding the assignment $p^i:=b$ if $p^i$ is in
  $P^i_j$, there exists an inter execution constraint of the form
  $p^i\Leftrightarrow p^o$ in $G_{\mbox{\scriptsize EC}}$, $p^o$ is in
  $P^o_k$, and $p^o:=b$ is among the parallel assignments of
  $\mathit{tr}_j$; otherwise, $\mathit{tr}_j$ is returned unchanged,
  for $j,k=1,2$ and $j\neq k$,
\item $B=B_1\cup B_2\cup G_{\mbox{\scriptsize Auth}}$, 
\item $P^i = \{p\in (P^i_1\cup P^i_2) | p \mbox{ does not occur in }
  G_{\mbox{\scriptsize EC}}\}$,
\item $P^o = \{p\in (P^o_1\cup P^o_2) | p \mbox{ does not occur in }
  G_{\mbox{\scriptsize EC}}\}$,
\item $H^o = H^o_1\cup H^o_2$,  and
\item $C^i = \{c\in (C^i_1\cup C^i_2) | c \mbox{ does not occur in }
  G_{\mbox{\scriptsize Auth}}\}$.
\end{itemize}
The definition is well formed since $S$ is obviously a
security-sensitive transition system and $\mathit{Int}$ satisfies all
the structural constraints at page~\pageref{page:def-comp}.
\begin{example}
  \label{ex:trw-comp}
  Let us consider components $\mathcal{C}_1$ and $\mathcal{C}_{234}$
  of Example~\ref{ex:trw-ex}. We glue them together by using the
  following set $G=G_{\mbox{\scriptsize EC}}\cup G_{\mbox{\scriptsize
      Auth}}$ of gluing assertions where $G_{\mbox{\scriptsize EC}} :=
  \{ p1_{1}\Leftrightarrow p0_{234}\}$ and $G_{\mbox{\scriptsize
      Auth}} := \{ \forall u.c^i_{t2}(u)\Leftrightarrow \neg
  h_{t1}(u), \forall u.c^i_{t4}(u)\Leftrightarrow \neg h_{t1}(u)\}$.
  The inter execution constraint in $G_{\mbox{\scriptsize EC}}$
  corresponds to the dashed arrow connecting $p1$ in component
  $\mathcal{C}_1$ ($p1_{1}$) to $p0$ in component $\mathcal{C}_{234}$
  ($p0_{234}$) in Figure~\ref{fig:wf-pn}.  The always constraints in
  $G_{\mbox{\scriptsize Auth}}$ formalize the dashed lines linking
  task $t1$ of component $\mathcal{C}_1$ to tasks $t2$ and $t4$ of
  component $\mathcal{C}_{234}$.  The component obtained by gluing
  $\mathcal{C}_1$ and $\mathcal{C}_{234}$ together with $G$ (in
  symbols, $\mathcal{C}_1\oplus_G \mathcal{C}_{234}$) is such that
  \begin{itemize}
  \item its set of transitions contains all transitions in
    $\mathit{Tr}_{234}$ plus the transition in $\mathit{Tr}_{1}$
    modified to take into account the inter execution constraint in
    $G_{\mbox{\scriptsize EC}}$, i.e.
    \begin{eqnarray*}
      t1(u) : p0_{1}\wedge \neg d_{t1} \wedge a_{t1}(u) \to p0_{1},p1_{1},d_{t1},h_{t1}(u):=F,T,T,T||p0_{234}:=T
    \end{eqnarray*}
    ensuring that when the token is put in $p1_{1}$ it is also put in
    $p0_{234}$ (in this way, we can specify how the control flow is
    transferred from $\mathcal{C}_1$ to $\mathcal{C}_{234}$);
  \item its set of always constraints contains all the constraints in
    $B_1$ and $B_{234}$ plus those in $G_{\mbox{\scriptsize Auth}}$ so
    that the SoD constraints between task $t1$ in $\mathcal{C}_1$ and
    tasks $t2$ and $t4$ in $\mathcal{C}_{234}$ are added;
  \item if its interface is $(A,P^i,P^o, H^o,C^i)$, then
    $P^i:=\{p1_1\}$ since $p0_{234}$ occurs in $G_{\mbox{\scriptsize
      EC}}$, $P^0:=\{p7_{234}\}$ since $p1_{1}$ occurs in
    $G_{\mbox{\scriptsize EC}}$, and $C^i:=\emptyset$ since both
    $c^i_{t2}$ and $c^i_{t4}$ occur in $G_{\mbox{\scriptsize Auth}}$.
  \end{itemize}
  Notice that $\mathcal{C}_1\oplus_G \mathcal{C}_{234}$ can be
  combined with $\mathcal{C}_5$ so as to form a component
  corresponding to the TRW in Figure~\ref{fig:wf-ex}.  This is
  possible by considering the following set $G'=G'_{\mbox{\scriptsize
      EC}}\cup G'_{\mbox{\scriptsize Auth}}$ of gluing assertions
  where $G'_{\mbox{\scriptsize EC}} := \{ p7_{234}\Leftrightarrow
  p0_{5}\}$ and $G'_{\mbox{\scriptsize Auth}} := \{ \forall
  u.c^i_{t5}(u)\Leftrightarrow \neg h_{t2}(u)\wedge \neg h_{t3}(u)\}$.
  The inter execution constraint in $G'_{\mbox{\scriptsize EC}}$
  corresponds to the dashed arrow connecting $p7$ in component
  $\mathcal{C}_{234}$ ($p7_{234}$) to $p0$ in component
  $\mathcal{C}_{5}$ ($p0_{5}$) in Figure~\ref{fig:wf-pn}.  The always
  constraint in $G_{\mbox{\scriptsize Auth}}$ formalizes the dashed
  lines linking task $t5$ of $\mathcal{C}_5$ with tasks $t2$ and $t3$
  of $\mathcal{C}_{234}$.\qed
\end{example}

We now illustrate the computation of the wlp with respect to the
transitions of a composed component by means of an example.
\begin{example}
  \label{eq:ex-border-effects}
  Let us consider $(\mathcal{C}_1\oplus_G
  \mathcal{C}_{234})\oplus_{G'} \mathcal{C}_5$ of
  Example~\ref{ex:trw-comp} and the predicates
  \begin{eqnarray*}
    K_1&:=& \neg p0_1\wedge d_{t1} \quad
    K_5:= \neg p0_5\wedge  p1_5\wedge d_{t5} \\
    K_{234}&:=& \bigwedge_{i=0, ..., 6} \neg pi_{234} \wedge 
              d_{t2} \wedge  d_{t3} \wedge  d_{t4}
             \wedge  d_{s_{234}} \wedge  d_{j_{234}}
  \end{eqnarray*}
  whose conjunction $K$ characterizes the final states of TRW,
  i.e.\ those situations in which all tasks have been executed and
  there is just one token in place $p1_5$ (notice that $K_{234}$ does
  not mention $p7_{234}$ whose value is implied by $K_5$ and the inter
  execution constraints $p7_{234}\Leftrightarrow p0_{5}$ and similarly
  $K_1$ does not mention $p1_{1}$ whose value is implied by $K_{234}$ and
  the inter execution constraint in $p1_{1}\Leftrightarrow p0_{234}$).

  Now we compute $\mathsf{wlp}(t5, B, K)$ where $B$ is the union of
  $B_1$, $B_{234}$, $B_{5}$, $G_{\mbox{\scriptsize Auth}}$, and
  $G'_{\mbox{\scriptsize Auth}}$ given in Example~\ref{ex:trw-comp} by
  using (\ref{eq:wlp-def}):
  \begin{eqnarray}
    \label{eq:ex-formula}
    K_1\wedge K_{234} \wedge 
    (p0_{5}\wedge \neg d_{t5} \wedge a_{t5}(u)\wedge 
    \neg h_{t2}(u)\wedge \neg h_{t3}(u)) \, .
  \end{eqnarray}
  Notice how $K_1$ and $K_{234}$ have not been modified since the
  parallel updates of $t5$ do not mention any of the state variables
  in $\mathcal{C}_1$ and $\mathcal{C}_{234}$ but only those of
  $\mathcal{C}_5$, namely $p0_{5}$ and $d_{t5}$.  

  To illustrate how the computation of wlp takes into account the
  transfer of the control flow from one component to another, let us
  compute the wlp of (\ref{eq:ex-formula}) with respect to transition
  $j$ in component $\mathcal{C}_{234}$.  According to the definition
  of composition of components, transition $j_{234}$ becomes
  \begin{eqnarray*} 
    j^*_{234} &:& p4_{234}\wedge p5_{234}\wedge p6_{234}\wedge \neg d_{j_{234}} \\
    &&\to p4_{234}, p5_{234},p6_{234}, p7_{234}, d_{j_{234}}:= F,F,F,T,T||p0_{5}:=T \, .
  \end{eqnarray*}
  Notice the added assignment $p0_{5}:=T$ to take into account the
  inter execution constraint in $G'_{\mbox{\scriptsize EC}}$ (see
  Example~\ref{ex:trw-comp}) ensuring that when the token is put in
  $p7_{234}$, it is also put in $p0_{5}$.  By using
  (\ref{eq:wlp-def}), we have that $\mathsf{wlp}(j^*_{234}, B,
  (\ref{eq:ex-formula}))$ is
  \begin{eqnarray}
    \label{eq:ex-formula-bis}
    K_1\wedge 
    \left[
    \begin{array}{l}
      \neg p0_{234} \wedge \neg p1_{234} \wedge \neg p2_{234} \wedge \\ \neg p3_{234} \wedge 
      p4_{234} \wedge p5_{234} \wedge p6_{234} \wedge \\ \neg d_{t2} \wedge  \neg d_{t3} \wedge \neg d_{t4} \wedge  \neg d_{s_{234}} \wedge  d_{j_{234}} 
      \end{array}
      \right] \wedge 
      \left(
      \begin{array}{l}
        \neg d_{t5} \wedge a_{t5}(u)\wedge \\ \neg
        h_{t2}(u)\wedge \neg h_{t3}(u) 
      \end{array}
      \right) \, .
  \end{eqnarray}
  Notice how $K_1$ is left unmodified since it describes the state of
  component $\mathcal{C}_1$ and no gluing assertions involve state
  variables of $\mathcal{C}_1$ and those in the update of $j_{234}^*$,
  $K_{234}$ instead is modified substantially (see the predicate in
  square brackets) since $j_{234}$ is a transition of component
  $\mathcal{C}_{234}$, while the remaining part of
  (\ref{eq:ex-formula-bis}) is almost identical to the formula between
  parentheses in (\ref{eq:ex-formula}) except for the deletion of
  $p0_{5}$ because of the additional assignment $p0_{5}:=T$ in
  $j^*_{234}$, introduced to take into account the inter execution
  constraint in $G'_{\mbox{\scriptsize Auth}}$.

  An alternative way of computing $\mathsf{wlp}(j^*_{234}, B,
  (\ref{eq:ex-formula}))$ is the following.  Observe that the value of
  $p7_{234}$ is fixed to $T$ because of the inter execution constraint
  $p0_{5}\Leftrightarrow p7_{234}$ in $G_{\mbox{\scriptsize Auth}}$
  and the fact that (\ref{eq:ex-formula}) implies that $p0_5$ is $T$.
  Thus, we can consider the predicate $K_{234}\wedge p7_{234}$ and
  then compute $\mathsf{wlp}(j_{234}, B_{234}, K_{234}\wedge
  p7_{234})$ which is the predicate in square brackets of
  (\ref{eq:ex-formula-bis}).  By taking the conjunction of this
  formula with $K_1$ and the predicate obtained by deleting $p0_5$
  from $\mathsf{wlp}(t5, B_5\cup G_{\mbox{\scriptsize Auth}}, K_5)$ in
  which we delete $p0_5$ (because (\ref{eq:ex-formula}) implies that
  $p0_5$ is $T$) thereby obtaining the predicate between parentheses
  in (\ref{eq:ex-formula}), we derive (\ref{eq:ex-formula-bis}) as
  before.\qed
\end{example}
The last paragraph of the example suggests a modular approach to
computing wlp's.  It is indeed possible to generalize the process
described above and derive a modularity result for computing the wlp
of a complex component by using the wlp's of its components by taking
into account the gluing assertions.  We do not do this here because it
is not central to the applications of the notion of component
discussed in Section~\ref{sec:applications} below.  

\begin{theorem}
  \label{thm:oplus-props}
  Let $(S_k, \mathit{Int}_k)$ be a symbolic security-sensitive
  component for $k=1,2,3$, $G_{1,2}$ be a set of gluing assertions
  over $\mathit{Int}_1$ and $\mathit{Int}_2$, and $G_{2,3}$ be a set
  of gluing assertions over $\mathit{Int}_2$ and $\mathit{Int}_3$.
  Then,
  \begin{description}
  \item[Commutativity:] $(S_1,\mathit{Int}_1)\oplus_{G_{1,2}}
    (S_2,\mathit{Int}_2)= (S_2,\mathit{Int}_2)\oplus_{G_{1,2}}
    (S_1,\mathit{Int}_1)$ and
  \item[Associativity:] $((S_1,\mathit{Int}_1)\oplus_{G}
    (S_2,\mathit{Int}_2))\oplus_{G} (S_3,\mathit{Int}_3)=
    (S_1,\mathit{Int}_1)\oplus_{G} ((S_2,\mathit{Int}_2)\oplus_{G}
    (S_3,\mathit{Int}_3))$ for $G=G_{1,2}\cup G_{2,3}$.
  \end{description}
\end{theorem}
The proof is straightforward and based on the commutativity and
associativity of set union.  Notice that the associativity property
above is expressed by taking into account the union of the gluing
assertions over the interfaces of the reusable systems being combined.
\begin{example}
  Recall the components of Example~\ref{ex:trw-comp}.  Because of
  Theorem~\ref{thm:oplus-props}, we have that the TRW can be expressed
  as
  $\mathcal{C}_1\oplus_{G''}\mathcal{C}_{234}\oplus_{G''}
  \mathcal{C}_5$ for $G''=G\cup G'$ where $G,G'$ have been
  defined in Example~\ref{ex:trw-comp}.
  
  Notice that, despite the commutativity of the operator $\oplus$, the
  task in $\mathcal{C}_1$ will always be executed before all tasks in
  components $\mathcal{C}_{234}$ and $\mathcal{C}_5$ because of the
  gluing assertions in $G''$.  Thus, the component
  $\mathcal{C}_{234}\oplus_{G''}\mathcal{C}_1 \oplus_{G''}
  \mathcal{C}_{5}$ obtained by considering the components in a
  different order is equivalent to TRW.\qed
\end{example}
Appendix~\ref{app:patterns} shows how standard composition patterns
available in the literature for workflows can be expressed by using
the notion of components and the composition operator $\oplus$
introduced above.

\section{Applications}
\label{sec:applications}

We present two applications of security-sensitive components and their
modular combination which are made possible by the same modularity
result about the synthesis of run-time monitors for the WSP.

In~\cite{bertolissi2015}, we have shown how to automatically derive a
monitor capable of solving the run-time version of the Workflow
Satisfiability Problem (WSP)~\cite{crampton} of a security-sensitive
transition system.  As already discussed in the paragraph
``\textbf{Semantics of a security-sensitive component}'' in
Section~\ref{sec:reusable}, the notion of security-sensitive
transition system introduced here and that in~\cite{bertolissi2015}
are equivalent.  In particular, given a security-sensitive transition
system $((P,D,A,H,C),\mathit{Tr},B)$ we can derive an equivalent
security-sensitive transition system of the form
$((P,D,A',H,\emptyset),\{[[\mathit{tr}]]_B|\mathit{tr}\in
B\},\emptyset)$, which is precisely a security-sensitive transition
system of~\cite{bertolissi2015}, where $A'$ contains the variables in
$A$ and those in $C$ which are not mentioned in $B$.  Let
$\mathcal{RM}$ be the procedure which takes as input a
security-sensitive transition system $S=((P,D,A,H,C),\mathit{Tr},B)$,
applies the transformation above, and then the procedure for the
synthesis of run-time monitors described in~\cite{bertolissi2015},
which returns a Datalog~\cite{datalog} program $\mathcal{RM}(S)$
defining a predicate $\mathit{can\_do}(u,t)$ such that user $u$
can execute task $t$ and the workflow can successfully terminate iff
$\mathit{can\_do}(u,t)$ is a logical consequence (in the sense of
Datalog) of $\mathcal{RM}(S)\cup \mathcal{P}\cup \mathcal{H}$ (in
symbols $\mathcal{RM}(S), \mathcal{P}, \mathcal{H} \models
\mathit{can\_do}(u,t)$), where $\mathcal{P}$ is a Datalog program
defining the meaning of the predicates in $A$ (i.e.\ the authorization
policy) and $\mathcal{H}$ is a set of \emph{history facts} of the form
$h_t(u)$, recording the fact that user $u$ has executed task
$t$.\footnote{There is an established line of research (see,
  e.g.,~\cite{constraintdatalog}) that has used (variants of) Datalog
  to express authorization policies.}

We now show how to reuse $\mathcal{RM}$ for the modular construction
of run-time monitors for the WSP, i.e.\ we build a monitor for a
composite component by combining those for their constituent
components.  Let $G=G_{\mbox{\scriptsize EC}}\cup G_{\mbox{\scriptsize
    Auth}}$ be a set of gluing assertions where $G_{\mbox{\scriptsize
    EC}}$ is a set of {inter execution constraints} and
$G_{\mbox{\scriptsize Auth}}$ a set of always constraints over an
interface $(A,P^i,P^o$, $H^o,C^i)$, then $\langle{G}\rangle :=
\langle{G_{\mbox{\scriptsize EC}}}\rangle\cup
\langle{G_{\mbox{\scriptsize Auth}}}\rangle$, where $
\langle{G_{\mbox{\scriptsize EC}}}\rangle := \{p^i\leftarrow p^o |
p^i\Leftrightarrow p^o\in G_{\mbox{\scriptsize EC}} \}$ and
$\langle{G_{\mbox{\scriptsize Auth}}}\rangle := \{c^i(u)\leftarrow
\mathit{hst}^i(u) | \forall u.c^i(u)\Leftrightarrow
\mathit{hst}^i(u)\in G_{\mbox{\scriptsize Auth}} \}$.  Intuitively,
the shape of the Datalog clauses in $\langle{G_{\mbox{\scriptsize
      EC}}}\rangle$ models how the execution flow is transferred from
a component (that with an output place) to the other (that with an
input place).
\begin{theorem} 
  \label{thm:monitor-comp}
  Let $(S_k, \mathit{Int}_k)$ be a symbolic security-sensitive
  component, $S_k=((P_k,D_k,A_k,H_k,C_k),\mathit{Tr}_k,B_k)$,
  $\mathcal{H}_k$ is a set of (history) facts over $H_k$, and
  $\mathcal{P}_k$ a Datalog program (for the authorization policy)
  over $A_k$, for $k=1,2$. If $G$ is a set of gluing assertions over
  $\mathit{Int}_1$ and $\mathit{Int}_2$, then
  $\mathcal{RM}(S),\mathcal{H}_1,\mathcal{H}_2,\mathcal{P}_1,\mathcal{P}_2
  \models can\_do(u,t)$ iff
  $
    \mathcal{RM}(S_1), \mathcal{H}_1,\mathcal{P}_1, 
    \langle{G}\rangle,
    \mathcal{RM}(S_2), \mathcal{H}_2,\mathcal{P}_2 \models can\_do(u,t) 
  $
  , where $(S,Int) = (S_1,Int_1)\oplus_{G} (S_2,Int_2)$.
\end{theorem}

The idea underlying the proof of this theorem is that the monitors for
the components are computed by considering any possible values for the
variables in their interfaces.  The additional constraints in the
gluing assertions simply consider a sub-set of all these values by
specifying how the execution flow goes from one component to the other
and how the authorization constraints across components further
constrain the possible executions of a component depending on which
users have executed certain tasks in the other.

As anticipated above, Theorem~\ref{thm:monitor-comp} paves the way to
two applications which are discussed more in detail in the following.

\noindent \textbf{Scalability of the Synthesis of Run-Time Monitors.}
It is possible to decompose large workflows into smaller components by
using pre-existing techniques (see, e.g.,~\cite{reijers2011}),
generate monitors for each module and glue them, allowing us to solve
the WSP for very large workflows, which would be otherwise intractable
due to state space explosion as shown in~\cite{bertolissi2015}.  The
main obstacle to the monitor synthesis is the state space explosion
caused by the need of computing all the possible interleavings of task
executions and the execution of these tasks by the users.
Theorem~\ref{thm:monitor-comp} allows for splitting a large
security-sensitive workflow into smaller components, allowing one to
synthesize the monitors for such components with smaller state spaces
and then glue them together in order to build the monitor for the
composed component.  

To show the practical scalability of this approach, we have performed
a set of experiments with the random workflow generator
from~\cite{bertolissi2015}, which is capable of generating random
security-sensitive workflows with an arbitrary number of tasks and
composing them sequentially.  For the experiments, we have generated
components with a fixed size of 5 tasks and a varying number of
constraints. The number of constraints is specified as a percentage
(5\%, 10\% and 20\%) of the number of tasks in each component for
intra-component constraints and as a percentage of the total number of
tasks for inter-component constraints. Thus, in the configurations 5\%
and 10\% there are no intra-component constraints, while in the
configuration 20\% there is one for each component; for a workflow
with 100 tasks, there are 5 inter-component constraints in the
configuration 5\%, 10 in the configuration 10\% and 20 in the
configuration 20\%. The experiments have been conducted on a MacBook
Air 2014 with a 1.3GHz dual-core Intel Core i5 processor and 8GB of
RAM running MAC OS X 10.10.2.
\begin{figure}[t]
  \centering
  \includegraphics[scale=.25]{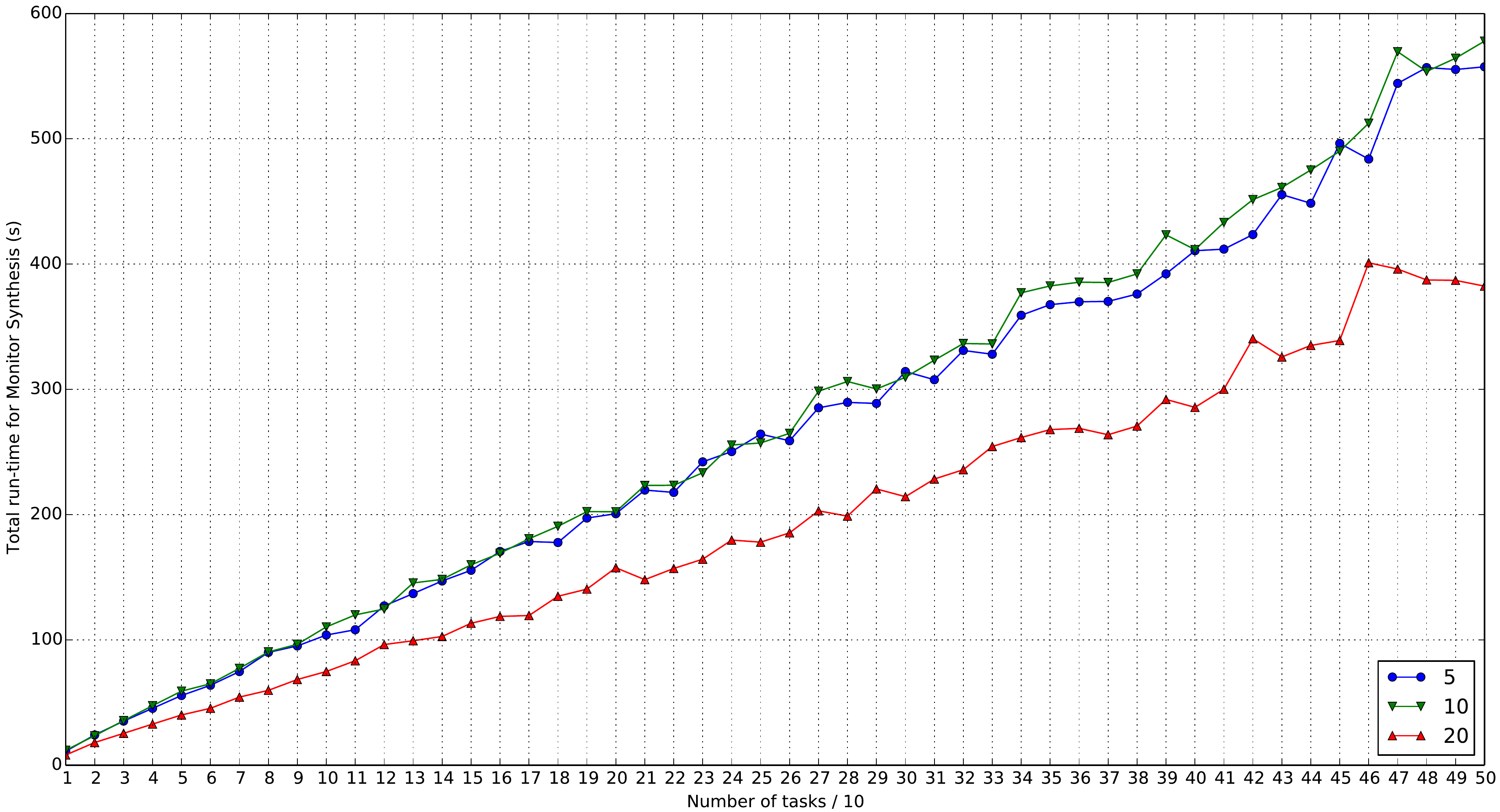}
  \caption{\label{fig:experiments} Time taken to synthesize a monitor varying 
with the number of tasks}
\end{figure}
The results are shown in Figure~\ref{fig:experiments}, in which the
x-axis contains the total number of tasks in a workflow divided by 10
(the total number of components is the number in the x axis times 2)
and the y-axis shows the total time in seconds taken by the monitor
synthesis procedure $\mathcal{RM}$ of~\cite{bertolissi2015}. Each data
point is taken as the average of running $\mathcal{RM}$ 5 times for
each configuration.  Figure~\ref{fig:experiments} suggests a linear
(instead of the expected exponential!) behavior with respect to the
number of tasks on this set of synthetic benchmarks.

\noindent \textbf{A Tool for the Design and Reuse of Components and
  Monitors}.  Recent practices in business process management have
emphasized the use of business process repositories~\cite{yan2012} in
order to promote process reuse and more quickly address the rapidly
evolving requirements on business process.
Theorem~\ref{thm:monitor-comp} supports not only the creation of
repositories containing reusable business processes in the form of
security-sensitive workflows but also associating with them run-time
monitors that can be modularly combined to create more complex
monitors for composed components.  These are important features in the
context of industrial applications of business processes, 
as they support reuse of existing technologies (editors and
repositories of business processes) and augment them by monitor
synthesis capabilities that make the synthesis automatic, scalable (as
shown by the experiments above), and transparent to the final user
(the procedure $\mathcal{RM}$ is fully automated).
\begin{figure}[t]
\centering
\includegraphics[scale=.5]{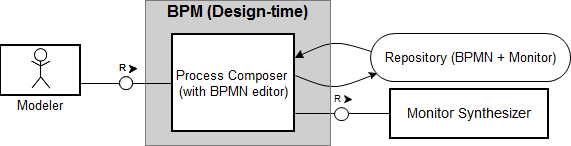}
\caption{\label{fig:implementation} Architecture of a business process
  design tool integrated with a repository of models and monitors}
\end{figure}
Figure~\ref{fig:implementation} outlines the high-level architecture
of a tool exploiting the ideas discussed here.  Rectangles represent
components, ovals represent storage systems, R-labeled links represent
request/response communication channels between components (where the
direction of the arrow states the direction of the request), and
arrows represent access to storages.  The \emph{BPM} (Business Process
Management) component represents any existing solution including a
modeling environment for BPMN-based business processes. The
\emph{Process Composer} sub-component is the modeling environment
offering a BPMN editor. Examples of BPM systems are IBM Business
Process
Manager\footnote{http://www-03.ibm.com/software/products/en/business-process-manager-family},
SAP Netweaver BPM\footnote{http://scn.sap.com/docs/DOC-27944},
and Signavio Process
Editor\footnote{http://www.signavio.com/products/process-editor/}.
The \emph{Monitor Synthesizer} component implements the procedure
$\mathcal{RM}$ described above to compute (modular) monitors for
workflow components and their composition modeled in the process
composer.  The \emph{Repository} component represents a storage system
for workflow models together with the monitor synthesized by the
(modular) monitor synthesizer. Note that such repository may be part
of the BPM solution (as in, e.g., IBM Business Process Manager) or
remotely located (e.g., Apromore~\cite{larosa2011}).  The
\emph{modeler} interacts with the process composer with a
request/response relation.
The same relation exists between the process composer and the monitor
synthesizer to request the synthesis of a run-time monitor for the
BPMN model under specification. The process composer can
store/retrieve BPMN models together with the synthesized monitors
to/from the repository.
\begin{example}
  Let us recall the situation in Example~\ref{ex:trw-ex}.  The tool in
  Figure~\ref{fig:implementation} allows us to re-use components
  $\mathcal{C}_1$ and $\mathcal{C}_5$ in both the specification of TRW
  and MDW as shown in Figure~\ref{fig:wf-pn}.  Additionally, the
  capability of storing automatically synthesized run-time monitors in
  the repository associated to the components permits their re-use in
  different business processes thanks to the modularity result in
  Theorem~\ref{thm:monitor-comp}.\qed
\end{example}
The business process modeling (process composer in
Figure~\ref{fig:implementation}) and repository components in the
proposed architecture are also part of common reference architectures,
e.g.,~\cite{weske2007}. The monitor synthesizer and the extension of
the repository to store monitors are unique contributions of this
paper.  Whenever a business \emph{modeler} uses the process composer,
he/she can import models with their associated monitors from the
repository, combine the models with new or pre-existing models and
export the resulting complex component back to the repository, storing
the process together with its monitor. Notice that the monitor
synthesis of the various components can be done, when necessary, while
the editing is progressing, thereby optimizing the waiting time for
the monitor.

So far, we have implemented the (modular) monitor synthesizer as a
command-line tool and not yet integrated it with a modeling
environment. We intend to do so using the extensible Signavio Core
Components\footnote{\url{https://code.google.com/p/signavio-core-components/}}
editor and a repository structure like Apromore~\cite{larosa2011},
which is already integrated with the editor and supports BPMN 2.0
models for processes. The repository must be extended to store parametric 
Datalog monitors associated with BPMN.
We believe that, since all the steps are automated, the graphical
integration will provide a very simple to use, push-button approach
for modelers to modularly and efficiently derive precise run-time
monitors for business processes that can be later securely deployed.

\section{Discussion}
\label{sec:conclusion}

We have described and formalized a modular approach for the synthesis
of run-time monitors for reusable security-sensitive workflows.  We
have shown the scalability of modular monitor synthesis by means of
experiments.
We have also discussed the initial implementation of a tool
integrating an editor with a repository of business processes extended
with the capability of storing associated run-time monitors so that
the modular synthesis of monitors can be exploited in business re-use.

Reuse in Business Process Management has been an intense topic of
research and industrial application; see,
e.g.,~\cite{koschmider2014,freitas2009}.
Several works in the field of Petri net have investigated modularity;
see, e.g.,~\cite{oanea2007}.  To the best of our knowledge, none of
the works in these contexts addresses security issues as we do here.
The most closely related work is~\cite{bertolissi2015}, which is
extended here with the notion of modularity.

As future work, we intend to fully implement the architecture in
Figure~\ref{fig:implementation} by using available repositories, such
as Apromore~\cite{larosa2011}.  We also plan to perform extensive
experiements concerning process reuse on the business processes
available in the repositories.
		
\bibliographystyle{plain}
\bibliography{biblio}

\begin{thebibliography}{10}
\bibitem{koschmider2014}
A.~Koschmider, M.~Fellmann, A.~Schoknecht, and A.~Oberweis.
\newblock Analysis of process model reuse: Where are we now, where should we go
  from here?
\newblock {\em Decision Support Systems}, 66(0):9 -- 19, 2014.
\end{thebibliography}

\newpage
\appendix

\section{Composition patterns}
\label{app:patterns}

We show how the basic control patterns in workflow management (see,
e.g.,~\cite{aalst2003,leuxner2010}) can be expressed by the gluing
operator $\oplus$ introduced above.  We consider \emph{sequential} 
(when $n$ processes are executed one after the other), 
\emph{parallel} (when $n$ processes are executed in parallel), and 
\emph{alternative} composition (when only one out of $n$ processes
is executed).  For lack of space, we do not consider other composition
patterns (such as the \emph{hierarchical} one, when a task is refined
to a complex process) which can also be expressed in our approach by
using a bit of ingenuity.  To simplify the technical development
below, we describe each composition pattern using two components
$(S_1,\mathit{Int}_1)$ and $(S_2,\mathit{Int}_2)$; the generalization
to $n$ components is straightforward.  Additionally, again for the
sake of simplicity, assume that $P_j^i=\{p_j^i\}$ and $P_j^o=\{p_j^o\}$ in
$\mathit{Int}_j=(A_j,P_j^i,P_j^o,H_j^o,C_j^i)$ for $j=1,2$,
i.e.\ there is just one input and just one output place in both
components.  (Notice that this assumption is satisfied when
considering workflow nets---see, e.g.,~\cite{aalst2000}---which are a
particular class of Petri nets frequently used for modeling 
workflows.)

\noindent \emph{Sequential composition.}  Let us consider the
situation in which the process specified by component $S_1$ must be
executed before the process executed by component $S_2$.  To model
this with the gluing operator, it is sufficient to consider a set
$G=G_{\mbox{\scriptsize EC}}\cup G_{\mbox{\scriptsize Auth}}$ of
gluing assertions over $\mathit{Int}_1$ and $\mathit{Int}_2$ such that
$G_{\mbox{\scriptsize EC}}=\{ p_2^i\Leftrightarrow p_1^o \}$.  Notice
that $(S_1,\mathit{Int}_1)\oplus_{G}
(S_2,\mathit{Int}_2)=(S_2,\mathit{Int}_2)\oplus_{G}
(S_1,\mathit{Int}_1)$ by Theorem~\ref{thm:oplus-props} but because the
gluing assertion in $G_{\mbox{\scriptsize EC}}$ is
$p_2^i\Leftrightarrow p_1^o$, and not $p_2^o\Leftrightarrow p_1^i$,
the process specified by component $(S_1,\mathit{Int}_1)$ will always
be executed before that specified by $(S_2,\mathit{Int}_2)$ when
considering their composition.  

\begin{figure}[t]
\centering
\includegraphics[scale=0.65]{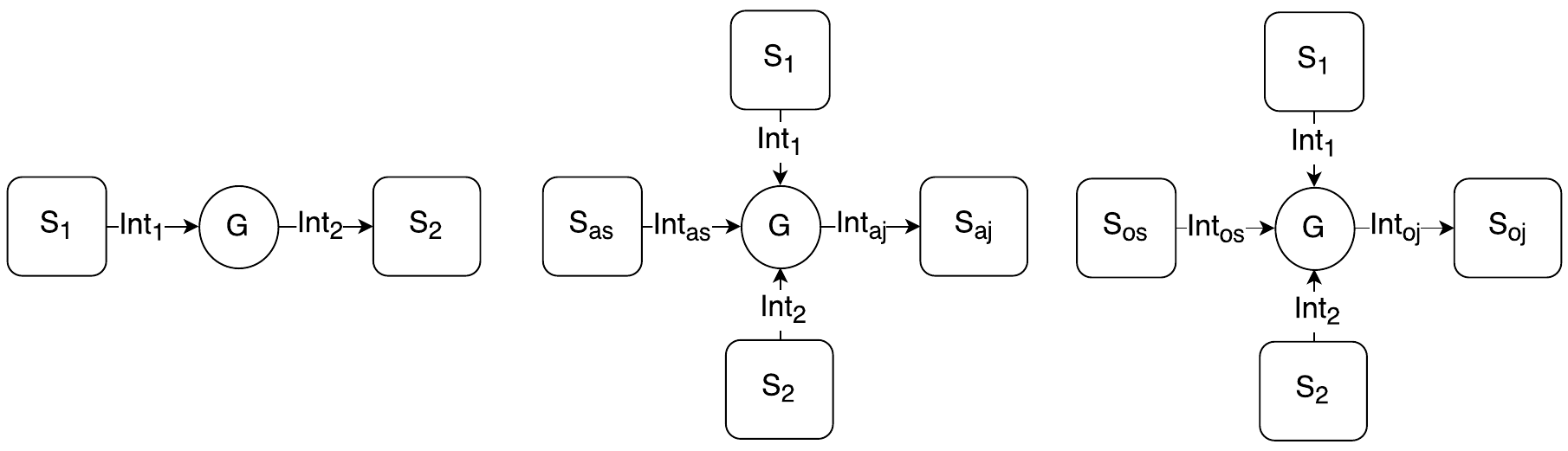}
\caption{\label{fig:patterns} Sequential (left), simultaneous (center) and 
alternative (left) composition}
\end{figure}

\noindent \emph{Parallel composition.}  Let us consider the situation
in which the processes specified by components $S_1$ and $S_2$ must be
executed in parallel.  To model this with the gluing operator, we need
to preliminarily introduce two other components, each containing a
single transition, one for splitting and one for joining the execution
flow.  Formally, we define
$C_*=((((P_*,D_*,\emptyset,\emptyset,\emptyset),\mathit{Tr}_*,\emptyset),(\emptyset,P^i_*,P^o_*,\emptyset,\emptyset)),
\mathit{Int}_*)$
, $P_{as}=\{p0_{as},p1_{as},p2_{as}\}$,
$D_{as}=\{d_{as}\}$, $P_{aj}=\{q0_{aj},q1_{aj},q2_{aj}\}$,
$D_{aj}=\{d_{aj}\}$,
\begin{eqnarray*}
  \mathit{Tr}_{as} &:=& \{ p0_{as}\wedge \neg d_{as} \rightarrow p0_{as},p1_{as},p2_s,d_{as}:=F,T,T,T\} \\
  \mathit{Tr}_{aj} &:=& \{ q0_{aj}\wedge q1_{aj}\wedge \neg d_{aj} \rightarrow  q0_{aj},q1_{aj},q2_{aj},d_{aj}:=F,F,T,T\} \, ,
\end{eqnarray*}
and $\mathit{Int}_*=(\emptyset,P_*^i,P_*^o,\emptyset,\emptyset)$ with
$P_{as}^i=\{p0_{as}\}$, $P_{as}^o=\{p1_{as},p2_{as}\}$, and
$P_{aj}^i=\{q0_{aj},q1_{aj}\}$, $P_{aj}^o=\{q2_{aj}\}$, where $*$
stands for $a$(nd) $s$(plit) or $a$(nd) $j$(oin).  At this point, it
is sufficient to consider a set $G=G_{\mbox{\scriptsize EC}}\cup
G_{\mbox{\scriptsize Auth}}$ of gluing assertions over
$\mathit{Int}_1$, $\mathit{Int}_2$, $\mathit{Int}_{as}$, and
$\mathit{Int}_{aj}$ (recall the associativity of the gluing operator
stated in Theorem~\ref{thm:oplus-props}) such that
$G_{\mbox{\scriptsize EC}}=\{ p1_{as}\Leftrightarrow p_1^i,
p2_{as}\Leftrightarrow p_2^i, p_1^o\Leftrightarrow q0_{aj},
p_2^o\Leftrightarrow q1_{aj}\}$. 

\noindent \emph{Alternative composition.}  Similarly to parallel
composition, we need to introduce also for this pattern two other
components, each containing two non-deterministic transitions to route
the execution flow in one of the two components
$(S_1,\mathit{Int}_1)$ or $(S_2,\mathit{Int}_2)$ instead of both as
above.  Formally, we define
$C_*=((((P_*,D_*,\emptyset,\emptyset,\emptyset),\mathit{Tr}_*,\emptyset),(\emptyset,P^i_*,P^o_*,\emptyset,\emptyset)),
\mathit{Int}_*)$ 
, $P_{os}=\{p0_{os},p1_{os},p2_{os}\}$,
$D_{os}=\{d_{os}\}$, $P_{oj}=\{q0_{oj},q1_{oj},q2_{oj}\}$,
$D_{oj}=\{d_{oj}\}$,
\begin{eqnarray*}
  \mathit{Tr}_{os} &:=& \left\{ 
  \begin{array}{l}
    p0_{os}\wedge \neg d_{os} \rightarrow p0_{os},p1_{os},p2_s,d_{os}:=F,T,F,T, \\
    p0_{os}\wedge \neg d_{os} \rightarrow p0_{os},p1_{os},p2_s,d_{os}:=F,F,T,T
  \end{array}
  \right\} \\
  \mathit{Tr}_{oj} &:=& \left\{ 
  \begin{array}{l} 
    q0_{oj}\wedge \neg d_{oj} \rightarrow  q0_{oj},q2_{oj},d_{oj}:=F,T,T, \\
    q1_{oj}\wedge \neg d_{oj} \rightarrow  q1_{oj},q2_{oj},d_{oj}:=F,T,T \\
  \end{array}
  \right\} \, ,
\end{eqnarray*}
and $\mathit{Int}_*=(\emptyset,P_*^i,P_*^o,\emptyset,\emptyset)$ with
$P_{os}^i=\{p0_{os}\}$, $P_{os}^o=\{p1_{os},p2_{os}\}$, and
$P_{oj}^i=\{q0_{oj},q1_{oj}\}$, $P_{oj}^o=\{q2_{oj}\}$, where $*$
stands for $o$(r) $s$(plit) or $o$(r) $j$(oin).  At this point, it is
sufficient to consider a set $G=G_{\mbox{\scriptsize EC}}\cup
G_{\mbox{\scriptsize Auth}}$ of gluing assertions over
$\mathit{Int}_1$, $\mathit{Int}_2$, $\mathit{Int}_{os}$, and
$\mathit{Int}_{oj}$ (recall the associativity of the gluing operator
stated in Theorem~\ref{thm:oplus-props}) such that
$G_{\mbox{\scriptsize EC}}=\{ p1_{os}\Leftrightarrow p_1^i,
p2_{os}\Leftrightarrow p_2^i, p_1^o\Leftrightarrow q0_{oj},
p_2^o\Leftrightarrow q1_{oj}\}$.

\end{document}